\documentclass[authoryear,review,12pt]{elsarticle}
\usepackage{graphicx}
\usepackage{amsmath,amsfonts}
\usepackage{lineno}
\linenumbers
\usepackage{subfigure}
\newcommand{\ud}{\mathrm{d}}

\makeatletter
\def\ps@pprintTitle{
\let\@oddhead\@empty
\let\@evenhead\@empty
\def\@oddfoot{\it \hfill\today}
\let\@evenfoot\@oddfoot}
\makeatother

\usepackage{endfloat} 
\newif\ifhavefigures
\havefigurestrue
\newif\ifnohavefigures 

\journal{\tiny Proceedings of the Royal Society B} 
\begin{document}
\begin{frontmatter}
  \title{Early Warning Signals and the Prosecutor's Fallacy}
  \author[cpb]{Carl Boettiger\corref{cor1}}
  \ead{cboettig@ucdavis.edu}
  \author[esp]{Alan Hastings}
  \cortext[cor1]{Corresponding author.}
  \address[cpb]{Center for Population Biology, 1 Shields Avenue, University of California, Davis, CA, 95616 United States.}
  \address[esp]{Department of Environmental Science and Policy, University of California, Davis} 


\begin{abstract}

Early warning signals have been proposed to forecast the possibility
of a critical transition, such as the eutrophication of a lake, the
collapse of a coral reef, or the end of a glacial period.  Because such
transitions often unfold on temporal and spatial scales that can be
difficult to approach by experimental manipulation, research has often
relied on historical observations as a source of natural experiments.
Here we examine a critical difference between selecting systems for
study based on the fact that we have observed a critical transition
and those systems for which we wish to forecast the approach of a
transition. This difference arises by conditionally selecting systems
known to experience a transition of some sort and failing to account
for the bias this introduces -- a statistical error often known as
the Prosecutor's Fallacy.  By analysing simulated systems that have
experienced transitions purely by chance, we reveal an elevated rate
of false positives in common warning signal statistics. We further
demonstrate a model-based approach that is less subject to this
bias than these more commonly used summary statistics.  We note that
experimental studies with replicates avoid this pitfall entirely.

\end{abstract}

 \end{frontmatter}

\section{Introduction}

\begin{quotation}
\noindent \emph{Mathematics \dots while assisting the trier of fact in the search of truth, must not cast a spell over him.}
-- California Supreme court, 1968.
\end{quotation}

\noindent In the case of \emph{People v. Collins} 1968, California Supreme
Court considered the evidence of an expert witness described by the
court as ``an instructor of mathematics at a state college'', which
concluded that the probability that a randomly selected individual
would match the description given by the victim would be less than 1 in
12 million~\citep{PeopleCollins1968}.  The prosecution had produced an
individual matching the prosecutor's detailed description, and convinced
by the mathematics, the lower courts had found him
guilty.

The prosecution has only observed that the probability of seeing the
evidence ($E$) they produced given a random innocent individual ($I$),
$P(E|I)$ is very small.  From this one cannot conclude that the individual
is indeed guilty, that is, that the probability the individual is innocent
given the evidence $P(I|E)$ is also very small. In a city with millions
of people, there might be several individuals who match the description
of the evidence.  Mathematically $P(E|I)$ need not equal $P(I|E)$,
instead, these expressions are related by Bayes theorem,

\begin{equation}
  P(E|I) = P(I|E) \frac{P(E)}{P(I)},
\end{equation} 

$P(E) \ll 1$ and $P(I)\approx 1$ , so $P(E|I) \approx P(I|E) P(E)$, and consequently we cannot conclude that $P(I|E) \ll 1$ from $P(E|I) \ll 1$.  
Realizing this mistake, the California Supreme Court reversed the decision, and
the case became a widely recognized example of the Prosecutor's
Fallacy~\citep{Thompson1987}.  Here we explore how a similar misconception
can arise from the use of historical data to evaluate methods for
detecting early warning signals of critical transitions.


Catastrophic transitions or tipping points, where a complex system
shifts suddenly from one state to another, have been implicated in
a wide array of ecological and global climate systems such as lake
ecosystems~\citep{Carpenter2011}, coral reefs~\citep{Mumby2007},
savannah~\citep{Kefi2007}, fisheries~\citep{Berkes2006}, and tropical
forests~\citep{Hirota2011}.  Recent research has begun to identify
statistical patterns commonly associated with these sudden catastrophic
transitions which could be used as an \emph{early warning sign} to
identify an approaching tipping point, which might provide managers time
to react to and avert an undesirable state shift~\citep{Scheffer2009, Lenton2011}.
An array of statistical patterns associated with tipping
point phenomena has been suggested for the detection of early warning
signals associated with such sudden transitions.  Two of the most commonly
used are a pattern of increasing variance~\citep{Carpenter2006} and a
pattern of increasing autocorrelation~\citep{vanNes2007}, which have
been tested in both experimental manipulation~\citep{Drake2010,
Carpenter2011, Veraart2011, Dai2012} and historical
observations~\citep{Livina2007,Dakos2008,Lenton2012,Ditlevsen2010,Guttal2008,Thompson2010}.

\subsection*{Testing patterns on historical data}

Historical examples of sudden transitions taken from the paleo-climate
record provide an important way to test and evaluate potential
leading indicator methods, and have been widely used for this purpose
\citep{Livina2007,Dakos2008,Lenton2012,Ditlevsen2010,Guttal2008,Thompson2010}.
Similarly, it has been suggested that data gathered from ecological
systems such as lakes that were monitored before they experienced sudden
eutrophication, or grasslands subjected to overgrazing, could contain
data that could help reveal when similar systems are approaching a
tipping point~\citep{Carpenter2011}.  

However, testing methods for early warning signals against historical
examples of transitions is susceptible to statistical mistakes that arise
from selecting data conditional on that data having already exhibited
a sudden transition.  A central tenant of early warning theory is that
the system in question is slowly approaching a tipping point that lies
some unknown distance away.  If nothing is done to remedy the situation,
this slow change will inevitably carry the system beyond the tipping
point, which introduces a sudden, rapid transition into an undesirable
state~\citep{Scheffer2009}. This process can be described mathematically
as a \emph{bifurcation}, in which a slowly changing parameter reaches
a critical value that causes the system stability to change.

Not all sudden transitions are caused by some ``guilty'' process slowly
driving the system over a tipping point -- the kind of process
that early warning signals are designed to detect.  Some systems may
experience such transitions purely by chance, leaving a stable state on
an extremely unlikely excursion that happens to stray to far from the
stable attractor~\citep[\emph{e.g.} ][consider this possibility in 
transitions that arise from analyzing historical climate record]{Ditlevsen2010, Lenton2011}. 
Like the evidence presented before the California Supreme
Court in 1968, the chance of observing such an ``innocent'' transition 
a priori may be very small, but when selected from a historical record
of many possible transitions, this possibility can no longer be ignored.

Figure 1 shows a schematic illustrating critical transitions under
each of these scenarios.  In the left panel, the system experiences a
bifurcation  and should contain an early warning signal.  In the right
panel, a similar-looking trajectory emerges from a simulation of a stable
system which should not contain a warning signal.  While the simulation of
the bifurcation scenario shown on the left produces a similar transition
every time, the transition shown on the right is somewhat less likely,
occurring in only 1\% of simulations.

\begin{figure}
  \begin{center}
      \ifhavefigures
        \includegraphics[width=\textwidth]{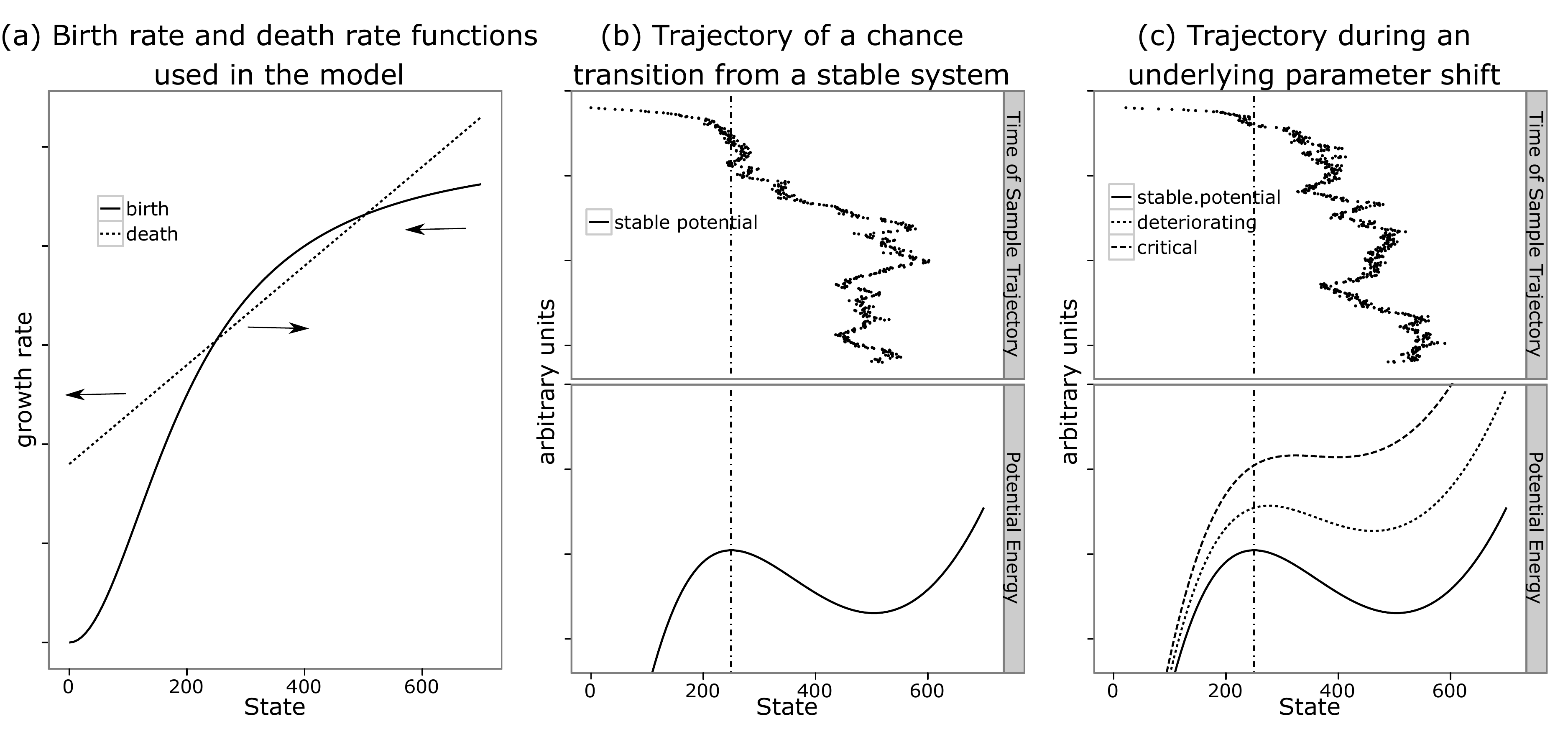}
      \fi
      \ifnohavefigures
        \hspace{\textwidth}
      \fi
      \label{fig:bifurcation}
  \end{center} \caption{\textbf{The Prosecutor's Fallacy}. 
\textbf{(a)} Plot of the model functions shown in Eq~\eqref{master}
with parameters $a=180$, $K=500$, $e=.5$, and $h=200$.  When the death
rate is higher than the birth rate, the system dynamics drive the state
(population size) to smaller values.  When birth rate is higher, the
system moves right, as indicated by the arrows.\textbf{(b)} The potential
energy is given by the negative integral of $b(n)-d(n)$, shown in the
lower plot.   The potential function gives an intuitive picture of
the stability of a system by imagining the curve as a surface on which
a ball is free to bounce across -- wells correspond to stable points
and peaks to unstable points. While most trajectories remain near the
stable well, some transition out merely by chance.  An example of such
a trajectory is shown in the top panel, in which time increases along
the vertical axis. Though initially oscillating around the stable state,
a chance excursion carries it beyond the Allee threshold (vertical dotted line).
Such chance trajectories can produce the statistical patterns as observed
in true critical transitions seen in panel \textbf{(c)}: Early warning
signals  are aimed at detecting systems which are slowly moving towards
a tipping point or bifurcation, illustrated in the successive curves
(deteriorating and critical). Top panel: An example trajectory from a
simulation under this process shows the state of the system as the
potential moves towards the bifurcation point. The original position of the Allee threshold
is shown by the vertical dotted line (though it moves slightly as the parameter changes).  
} \label{fig:1}
\end{figure}


\section{Methods and Results}
To investigate if early warning signals are vulnerable to this fallacy,
we simulate a system that is not driven towards a bifurcation such as
in Fig~ref{fig:1}(b).  This simulation approach allows us to determine whether
examining historical events is a valid way to test the utility of these
indicators.  We simulated 20,000 replicates of a stochastic individual-based
birth-death process with an Allee threshold~\citep{Courchamp2008}, which
arises from positive fitness effects at low densities.  Above the Allee
threshold the population returns to a positive equilibrium size, whereas
below the threshold the population decreases to zero. The model can be represented as a continuous time birth-death process where births and deaths are Poisson events which depend on the current density with rates given by

\begin{align}
    b(n) &= \frac{ K n^2}{n^2 + h^2}, \label{master} \\
    d(n) &= e n + a, 
\end{align}



a model with a linear death rate and density-dependent birth rate
that drives the Allee effect at low densities and limits growth at high
densities.  In this model $n$ indicates the discrete number of individuals
in the population, $K$ indicates a carrying capacity as set by a limiting
resource, $e$ a per-capita death rate (the $e$ scaling term in the birth
equation allows the carrying capacity $K$ to correspond to a positive
equilibrium point), $a$ an additional mortality imposed on the population
such as harvest, $h$ is a parameter controlling at what population size
the addition of more individuals switches from conferring a positive
benefit on growth from Allee interactions $n < h$ to a negative impact
on growth due to increased competition, $n > h$.  The key feature of
this model is the alternate stable states introduced by this effect;
other functional forms for Eq.~\eqref{master} could serve equally
well for these simulations~\citep[see \emph{e.g.}][]{Scheffer2001}.
Though this system can be forced through a bifurcation by increasing
the death rate, in these simulations all parameters are held constant
and no bifurcation occurs.  Consequently we do not anticipate an early
warning signal of an approaching bifurcation.

The simulation starts from the positive equilibrium population size.
Though the chance of a transition across the Allee threshold in any
given time step is small, given enough time this system will eventually
experience such a rare event driving the population extinct.  We ran
each replicate over 50,000 time units, sampling the system every 50
time units.  In this time window 266 of the 1,000 replicates experience
population collapse.  To keep the examples of comparable sample size,
we focus on a section of the data 500 time points prior to the system
approaching the transition.

To test whether selecting systems that have experienced
spontaneous transitions could bias the analysis towards false
positive detection of early warning signals, (the Prosecutor's
Fallacy) we selected replicates conditional on having collapsed
in the simulations.  We then selected a window around each system
that ended just before the collapse, while the population values
were still above the Allee threshold.  For each replicate, we
calculated the most common early warning indicators, variance and
autocorrelation~\citep[\emph{e.g.}][]{Carpenter2006,Dakos2008,Scheffer2009},
around a moving window equal to half the length of that time series.

\begin{figure}
  \begin{center}
    \ifhavefigures
    \includegraphics[width=6in]{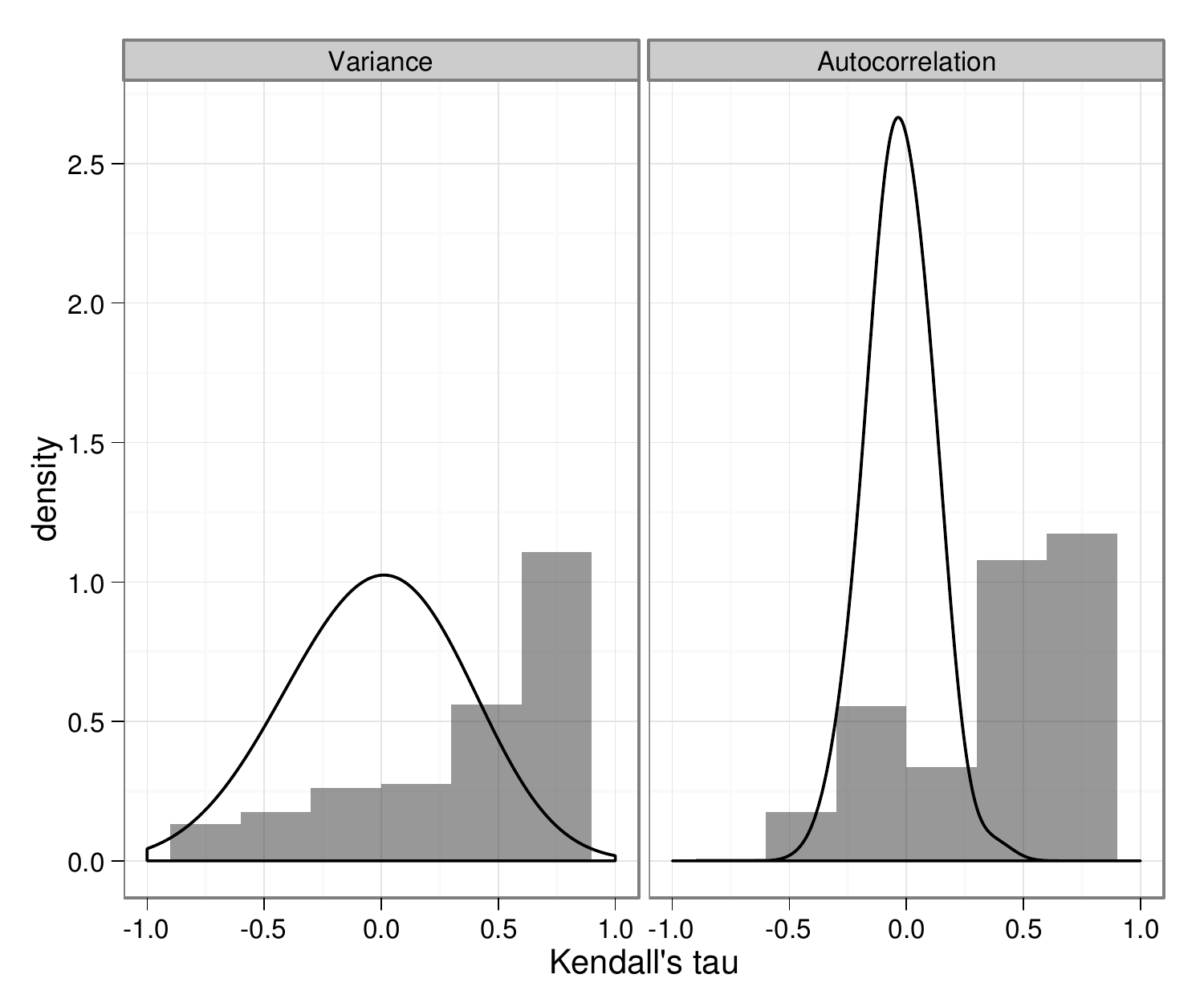}
    \fi
     \ifnohavefigures
     \hspace{6in}
    \fi
 \end{center}
  \caption{The distribution of the correlation statistic $\tau$ for two
  early warning indicators (variance, autocorrelation) on replicates
  conditionally selected for having collapsed by chance in simulations
  is shown in grey bars.  Solid lines indicate the estimated density of
  the statistic from a random sample of the simulations (not conditional
  on observing a transition). Positive values of $\tau$ correspond to
  a pattern of an indicator increasing with time; typically taken as
  evidence that a system is approaching a critical transition.  In these
  simulations, the pattern arises instead from the Prosecutor's fallacy
  of conditional selection.}
  \label{fig:indicator}
\end{figure}

To test for the presence of a warning signal in these indicators we
computed values of Kendall's $\tau$ for both indicators for each of
the 266 replicates.  Kendall's $\tau$ is a non-parametric measure
of rank correlation frequently used to identify an increasing trend
($\tau > 0 )$ in early warning signals~\citep{Dakos2008, Dakos2011},
defined as $\tau \def \tfrac{(\text{number of concordant pairs}) -
(\text{number of discordant pairs})}{\frac{1}{2} n (n-1) }$ in $n$
observations.  \footnote{A pair of observations $(x_i,y_i)$ and $(x_j,
y_j)$ are concordant if $x_i > x_j$ and $y_i > y_j$ or $x_i < x_j$ and
$y_i < y_j$ and discordant otherwise; equalities excepted.}  $\tau$ takes
values in $(-1, 1)$.  The distribution of $\tau$ values observed across
these replicates is shown in Figure~\ref{fig:indicator}.  We compare
the distribution of $\tau$ from all the simulations to the distribution
conditioned on experiencing a chance transition to the alternative
stable state.  To avoid an effect of sample size the time series are
all chosen to be the same length.

\begin{figure}
  \begin{center}
    \ifhavefigures
    \includegraphics[width=6in]{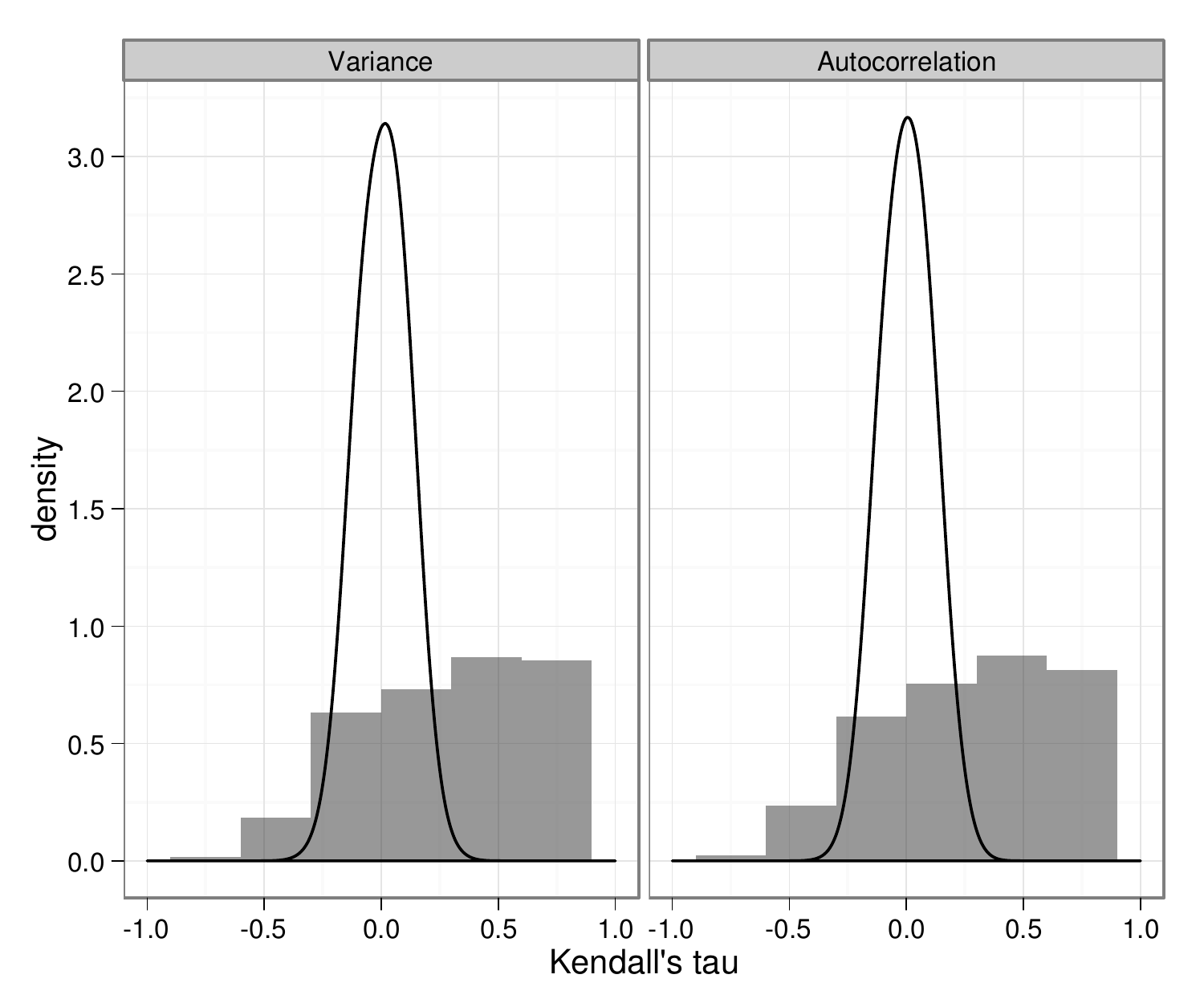}
    \fi
    \ifnohavefigures
     \hspace{6in}
    \fi
  \end{center}
  \caption{The identical analysis from Figure~\ref{fig:indicator} is shown for the model in Eq~\eqref{May1977} using parameters $r = 0.75$, $K=10$, $a=1.7$, $Q=3$, and $H=1$. A similar statistical bias, particularly towards positive values of $tau$ occurs in this model as well.}
  \label{fig:may}
\end{figure}

To demonstrate the effect we observe is not unique to models with Allee
effects, we provide an example of the effect arising in a discrete-time
model with two non-zero stable states adapted from~\citep{May1977},
\begin{equation}
X_{t+1} = X_t \exp\left( r \left(1 - \frac{ X_t }{ K } \right) - \frac{ a * X_t ^ {Q - 1} }{X_t ^ Q + H ^ Q} \right). \label{May1977}
\end{equation}
which combines a logistic growth model with a saturating predator response (See~\citet{May1977} for detailed discussion), shown in Figure~\ref{fig:may}.  Code to replicate the analysis can be found at \\ 
\verb|https://github.com/cboettig/earlywarning/tree/prosecutor/|.

For each of these replicates we also take a model-based
approach, estimating parameters for an approximate linear
model of the system approaching a saddle node bifurcation, as
described by~\citet{Boettiger2012b},

\begin{equation}
\ud X = \sqrt{ r_t } (\phi(r_t) - X_t)\ud t + \sigma\sqrt{\phi(r_t) } \ud B_t \label{LSN}
\end{equation}

In this model, the parameter $m$
describes the approach towards the saddle-node bifurcation.  Estimates
$m < 0 $ are expected in systems approaching a bifurcation, while for
stable systems $m$ should be approximately zero. None of the estimates
across the 266 simulations differed from zero in our study, hence the
model-based estimation shows no evidence of bias on data that has been
selected conditional on collapse.

\section{Discussion}

The attempts to detect early warning signs for critical transitions
are based on the concept of a deteriorating environment as embodied
in a changing parameter~\cite{Scheffer2009}, which is a different kind
of transition than one which is driven instead by stochasticity in an
environment which is otherwise constant and exhibiting no directional
change. When trying to use historical data to understand critical
transitions we often do not know which category, changing environment
or simply chance, an observed large change falls into.

We have shown here that systems which undergo rare sudden transitions
due to chance look statistically different from their counterparts that
do not, even though they are driven by the same stochastic process.
In particular, such conditionally selected examples are more likely to
show signs associated with an early warning of an approaching tipping
point, such as increasing variance or increasing autocorrelation,
as measured by Kendall's $\tau$.  This increases the risk of false
positives -- cases in which a warning signal being tested appears to
have successfully detected an underlying change in the system leading
to a tipping point, when in fact the example comes instead from a stable
system with no underlying change in parameters. Figure 2 shows that many
of the chance crashes show values of $\tau$ that are significantly larger
than those observed in the otherwise identical replicates that did not
experience a chance transition, thus ``detecting'' an underlying change
in the system dynamics that is not in fact present.

\subsection{Chance transitions are false positives for early warning signals}

It seems tempting to argue that this bias towards positive detection
in historical examples is not problematic -- each of these systems did
indeed collapse, so the increased probability of exhibiting warning
signals could be taken as a successful detection.  Unfortunately this is
not the case. At the moment the forecast is made, these systems are not
likely to transition, since they experience a strong pull towards the
original stable state.  A closer look at the patterns involved shows why
common indicators such as autocorrelation and variance can be misleading.

As the system gets farther from its stable point, it it more likely to
draw a random step that returns it towards the stable point. Despite
this, there is always some probability that it will move further still,
so systems that do cross the tipping point must do so rather quickly by
a string of events.  This pattern, clearly visible before the crashes
in each of the examples in Figure 1, produces a string of observations
that appear more highly autocorrelated (if we are sampling the system
frequently enough to catch the excursion at all) than we observe in the
rest of the fluctuations around the equilibrium.  Yet this autocorrelation
comes from a chance trajectory moving quickly \emph{away} from the
stable state, not from the critical slowing down pattern in the return
times to the stable state which precede a saddle-node bifurcation and
motivate the early warning signal.

This longer than expected excursion results in a higher than expected
variance in that window as well. Both variance and autocorrelation are
calculated using a moving window over the time-series, which allows
the method to pick out a pattern of change as the window moves along
the sequence. If this chance excursion that precedes the crash happens
to fill a significant part of the moving window, the resulting pattern
will tend to show an increase in autocorrelation or variance.  If the
chance excursion is relatively rapid compared to the frequency at which
the system is observed (spacing of the data) or the width of the moving
window, the excursion may not significantly alter the general pattern.
In this way, some of the events in which a crash is observed will
appear to present these statistical patterns of increased variance
or autocorrelation without being harbingers of approaching critical
transitions.

\subsection{ The truncation of observations }

If we had a complete knowledge of the system dynamics, then we could
eliminate the bias we observe here since the bias arises from the
transient branch of the trajectory that crosses the threshold, and
if the system were truncated at the minimum of the potential then the
effect we emphasize here would not appear.  But, it is not possible to
truncate the system in any practical application. The precise location of
the minimum of the potential which is the location of the deterministic
equilibrium is unknown. Moreover, under the hypothesis that the system is
approaching a critical transition, the location of the minimum potential
moves so it cannot easily be estimated by previous observations, (see
Figure 1c where the equilibrium point moves in the direction of the
transition). Thus it is neither practical nor desirable to suggest that
historical time series can be used by following a simple truncation rule
that avoids the branch of a trajectory crossing the threshold to another
basin of attraction.  Exactly where a particular study will choose
to truncate such a trajectory will necessarily be arbitrary without an
underlying model of the process.  Frequently this is done by removing
the very steep, monotonic branch of the trajectory expected once the
system crosses the unstable threshold.  Such an approach corresponds
with our choice of termination and produces the bias we discuss here.

The examples of Figure 1, though only single replicates, may be useful
in illustrating these issues.  Figure 1c, top panel shows a sample
trajectory of a system with a parameter shift, while 1b shows a trajectory
without a shift.  Both trajectories become more highly autocorrelated
and higher variance near the end of the time series (time increases
on the y axis in Figure 1).  The part of the time series following
the critical transition shows a fast and monotonic trajectory to the
unstable trajectory, and would usually be excluded by an analysis for
warning signals in advance of the transition.  No such clear pattern
exists prior to the transition in Figure 1b.   An alternative proposal
to terminate the trajectory in panel B earlier would also risk decreasing
the signal seen in panel c, and would be inconsistent with the application
of warning signals in the forecasting context, where there would be no
such truncation.

\subsection{Comparing to the model-based method}

In our numerical experiment, the model-based estimate of early warning
signals appears more robust than the summary statistics, producing
the same estimates on both the conditionally selected replicates as on
a random sample of the replicates.  This is a consequence of the more
rigid specifications that come with a model-based approach -- the pattern
expected is less general than any increase in variance or autocorrelation,
but instead must be one that matches its approximation of the saddle-node
bifurcation. This observation highlights the difference between the
pattern driving the false positive trends in increasing variance and
increasing autocorrelation and the pattern anticipated in the saddle-node
model. This should not however be taken as evidence that the model-based
approach is immune to the bias of the Prosecutor's Fallacy.

\subsection{Importance of experimental approaches} The problem we
highlight ultimately stems from the difficulty of having only a single
realization with which to examine a complex problem.  The only way
to deal with this problem embodied is through replication, as can
be done in an experimental system in laboratory manipulations such
as~\citet{Drake2010, Veraart2011, Dai2012} and at the scale of whole
lake ecosystems in~\citet{Carpenter2011}.  Experimental procedures avoid
the hazard of the Prosecutor's fallacy by generating a complete sample
of replicates, rather than selecting a subset of cases from some larger
historical sample.

 \section{Acknowledgments}
This research was supported by funding from NSF Grant EF 0742674 to AH
and a Computational Sciences Graduate Fellowship from the Department
of Energy grant DE-FG02-97ER25308 and NERSC Supercomputing grant
DE-AC02-05CH11231 to CB. The authors thank M. Baskett, T.A. Perkins
and N. Ross for helpful comments on earlier drafts of the manuscript,
and also P. Ditlevsen and an anonymous reviewer for their comments.

 \end{document}